\documentclass[]{aa}

\usepackage{graphicx}
\usepackage{epsfig}
\usepackage{natbib}
\usepackage{txfonts}
\usepackage{color,soul}
\usepackage{hyperref}

\definecolor{MyRed}{rgb}{0.9,0.0,0.0} 
\definecolor{MyLightRed}{rgb}{1.0,0.0,0.0} 
\definecolor{MyPink}{rgb}{1.0,0.08,0.45} 
\definecolor{MyDarkBlue}{rgb}{0,0.08,0.45} 
\definecolor{MyDarkGreen}{rgb}{0,0.5,0.0} 

\newcommand{\COBOLD}{{\tt CO$^5$BOLD}}
\newcommand{\Stagger}{{\tt Stagger}}
\newcommand{\MURAM}{{\tt MURaM}}
\newcommand{\BIFROST}{{\tt Bifrost}}
\newcommand{\LHD}{{\tt LHD}}
\newcommand{\MARCS}{{\tt MARCS}}
\newcommand{\Linfor}{{\tt Linfor3D}}
\newcommand{\NLTEIIID}{{\tt NLTE3D}}
\newcommand{\ATLAS}{{\tt ATLAS9}}

\newcommand{\PHOENIX}{{\tt PHOENIX}}

\newcommand{\moh}{\ensuremath{[\mathrm{M/H}]}}
\newcommand{\aoFe}{\ensuremath{\left[\mathrm{\alpha}/\mathrm{Fe}\right]}}

\newcommand{\Teff}{\ensuremath{T_{\mathrm{eff}}}}

\newcommand{\xtmean}[1]{\ensuremath{\left\langle #1\right\rangle}}
\newcommand{\mlp}{\ensuremath{\alpha_{\mathrm{MLT}}}}

\newcommand{\var}[1]{{\ensuremath{\sigma^2_{#1}}}}
\newcommand{\sig}[1]{{\ensuremath{\sigma_{#1}}}}

\newcommand{\beq}{\begin{equation}}
\newcommand{\eeq}{\end{equation}}
\newcommand{\Blam}{\ensuremath{B_\lambda}}

\newcommand{\intlam}{\ensuremath{\int_0^{\infty}\!d\lambda\,}}
\newcommand{\dT}{\ensuremath{\Delta T}}

\newcommand{\vartot}{\ensuremath{\frac{\var{T}}{T^2}}}

\newcommand{\eref}[1]{\mbox{(\ref{#1})}}

\newcommand{\calspec}{\emph{HST}~{\tt CALSPEC}}

\begin{document}

\title{Using the CIFIST grid of \COBOLD\ 3D model atmospheres to study the effects of stellar granulation on photometric colours.}
\subtitle{II. The role of convection accross the H--R diagram}

\author{
        A.~Ku\v{c}inskas\inst{1}
        \and
        J.~Klevas\inst{1}
        \and
        H.-G.~Ludwig\inst{2}
        \and
        P.~Bonifacio\inst{3}
                \and
                M.~Steffen\inst{3,4}
        \and
        E.~Caffau\inst{3}
       }

\offprints{A.~Ku\v{c}inskas}

\institute{
        Institute of Theoretical Physics and Astronomy, Vilnius University, A. Go\v {s}tauto 12, Vilnius LT-01108, Lithuania \\
        \email{arunas.kucinskas@tfai.vu.lt}
        \and
        ZAH Landessternwarte K\"{o}nigstuhl, D-69117 Heidelberg, Germany
        \and
                GEPI, Observatoire de Paris,  PSL Research University, CNRS,  Place Jules Janssen, 92195 Meudon, France
        \and
        Leibniz-Institut f\"ur Astrophysik Potsdam, An der Sternwarte 16, D-14482 Potsdam, Germany
}

\date{Received 11 December 2016 / Accepted 12 January 2017}

\abstract
{}
{We studied the influence of convection on the spectral energy distributions, photometric magnitudes, and colour indices of different types of stars across the H--R diagram.}
{The 3D hydrodynamical \COBOLD, averaged $\xtmean{\mbox{3D}}$, and 1D hydrostatic \LHD\ model atmospheres were used to compute spectral energy distributions of stars on the main sequence (MS), main sequence turn-off (TO), subgiant branch (SGB), and red giant branch (RGB), in each case at two different effective temperatures and two metallicities, $\moh=0.0$ and $-2.0$. Using the obtained spectral energy distributions, we calculated photometric magnitudes and colour indices in the broad-band Johnson-Cousins $UBVRI$ and 2MASS $JHK_{\rm s}$, and the medium-band Str\"{o}mgren $uvby$ photometric systems.}
{The 3D--1D differences in photometric magnitudes and colour indices are small in both photometric systems and typically do not exceed $\pm0.03$\,mag. Only in the case of the coolest giants located on the upper RGB are the differences in the $U$ and $u$ bands able reach  $\approx-0.2$\,mag at $\moh=0.0$ and  $\approx-0.1$\,mag at $\moh=-2.0$. Generally, the 3D--1D differences are largest in the blue-UV part of the spectrum and decrease towards longer wavelengths. They are also sensitive to the effective temperature and are significantly smaller in hotter stars. Metallicity also plays a role and leads to slightly larger 3D--1D differences at $\moh=0.0$. All these patterns are caused by a complex interplay between the radiation field, opacities, and horizontal temperature fluctuations that occur due to convective motions in stellar atmospheres. Although small, the 3D--1D differences in the magnitudes and colour indices are nevertheless comparable to or larger than typical photometric uncertainties and may therefore cause non-negligible systematic differences in the estimated effective  temperatures.
}
{}

\keywords{ Hydrodynamics -- Stars: atmospheres -- Techniques: photometric -- Techniques: spectroscopic}

\authorrunning{Ku\v{c}inskas et al.}
\titlerunning{Influence of stellar granulation on photometric colours}

\maketitle

\section{Introduction}

Our understanding of stars relies on the analysis of their spectroscopic and photometric properties, which are studied with the aid of stellar model atmospheres. Obviously, physical realism of the model atmospheres is one of the key factors that define the reliability of the obtained results. Until now, 1D hydrostatic model atmospheres were the most widely used  in studies of this kind. Amongst the advantages that make these models particularly attractive is that they combine sufficiently realistic physics with high computational speeds and the ease of use. Not surprisingly, a number of extensive 1D model atmosphere grids have been produced so far and have been widely used in many astrophysical contexts (e.g.  \ATLAS, \citealt[][]{CK03}; \MARCS, \citealt[][]{GEK08}; \PHOENIX, \citealt[][]{BH05}). Nevertheless, since these are 1D hydrostatic models, they cannot properly account for the multidimensional and time-dependent nature of phenomena taking place in stellar atmospheres, such as convection, pulsations, shock-wave activity.

A step beyond these limitations can be made by using state-of-the-art 3D hydrodynamical model atmospheres. Currently, several 3D hydrodynamical model atmosphere codes are available for computing realistic model atmospheres of different types of stars, for example \BIFROST\ \citep{GCH11}, \COBOLD\ \citep{FSL12}, \MURAM\ \citep{V03,VSS05}, and \Stagger\ \citep[][and references therein]{MCA13}. A quickly growing body of evidence shows that these models are indeed superior to their 1D counterparts in many aspects, including their ability to better reproduce observable properties of different types of stars: the Sun \citep{ANT00,CLS08,CLS11,SPC15}, late-type dwarfs \citep{GHB09,RAP09,BBL10}, subgiants \citep{GHB09,AAC15}, and red giants \citep{CAT07,RCL10,DKS13}. It is also important to note that the first grids of 3D hydrodynamical model atmospheres covering a range in $\Teff$, $\log g$, and $\moh$ are already available for use  (e.g. CIFIST grid produced using \COBOLD\ model atmospheres, \citealt{LCS09}; the grid computed with the \Stagger\ code, \citealt{MCA13}). All these important developments  form a solid basis for the wider use of 3D hydrodynamical models in stellar atmosphere studies.

Unfortunately, the influence of convection on the spectral energy distributions (SEDs), photometric magnitudes, and colour indices -- and thus the potential advantages of using 3D model atmospheres in this domain -- still remains largely unexplored. In our earlier studies \citep[][]{KHL05,KLC09}, we used 3D hydrodynamical \COBOLD\ model atmospheres to compute photometric magnitudes and colour indices of a red giant located close to the red giant branch (RGB) tip ($\Teff\approx3660$~K, $\log g=1.0$, $\moh=0.0$). We found that convection plays a non-negligible role in defining the SED, photometric magnitudes, and colour indices of this particular object, with noticeable 3D--1D differences in the colour indices (e.g. $\Delta (V-I) \approx 0.14, \Delta (V-K) \approx 0.20$) and, subsequently, differences in the effective temperature (determined using $\Teff$ -- colour calibrations) of up to $\sim60$~K. A similar investigation was also performed by \citet[][]{C09} for the Sun and a K-type dwarf ($\Teff=4780$~K, $\moh=0.0$). In this case, however, the 3D--1D differences in \Teff\ were smaller and did not exceed 10~K and 20~K, respectively. To our knowledge, these are the only studies where photometric magnitudes and colour indices of stars have been analysed using 3D hydrodynamical model atmospheres. As a consequence, the impact of convection on the spectrophotometric properties of stars in a wider atmospheric parameter space is still unknown.

We  therefore performed a detailed investigation of the influence of convection on the SEDs, photometric magnitudes, and colour indices. The obtained results are summarized in a series of two papers. In the first paper, \citet[][hereafter Paper~I]{BCL18}, we presented grids of photometric colour indices and 3D--1D corrections in several widely used photometric systems, computed using 3D hydrodynamical \COBOLD\ and 1D hydrostatic \LHD\ model atmospheres. These grids can be used to correct synthetic photometric colour indices for the 3D hydrodynamical effects and therefore can be applied to any type of study based on stellar photometry.

In the current study, which is the second paper in the series, we used a smaller set of 3D hydrodynamical \COBOLD\ and 1D hydrostatic \LHD\ model atmospheres ranging from the MS to the RGB tip at two different metallicities, $\moh=0.0$ and $-2.0$. Three types of model atmospheres were utilized to compute synthetic SEDs, photometric magnitudes, and colour indices: 3D hydrodynamical, averaged $\xtmean{\mbox{3D}}$, and 1D hydrostatic (see Sect.~\ref{sect:method-models} for details). We focused on a detailed comparison of the model predictions in order to understand the influence of convection on the spectrophotometric properties of stars across the H--R diagram.

The present paper is structured as follows. The methodology used to compute synthetic photometric magnitudes and colour indices is outlined in Sect.~\ref{sect:method}. The obtained results, including the computed SEDs of the Sun and Vega, and the SEDs and photometric magnitudes of the model atmospheres across the H--R diagram, as well as the influence of various model assumptions on the computed SEDs and magnitudes/colour indices, are presented and discussed in Sect.~\ref{sect:discussion}. Finally, the main results and conclusions are summarized in Sect.~\ref{sect:conclus}.

\begin{figure}[tb]
\centering
\includegraphics[width=9cm]{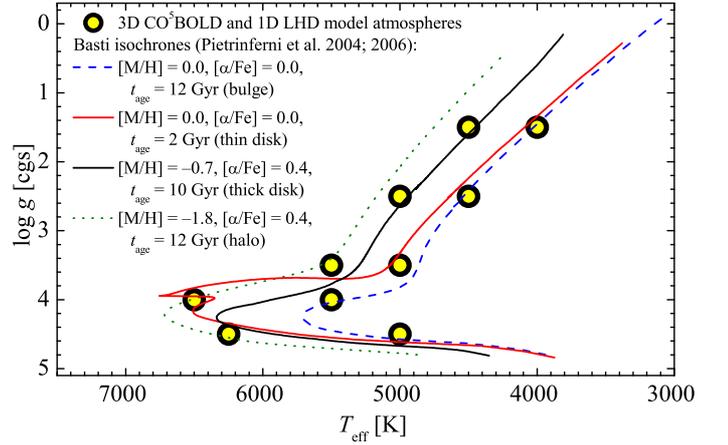}
\caption
{3D hydrodynamical \COBOLD\ and 1D hydrostatic \LHD\ model atmospheres used in this work, shown in the $\log g - \Teff$ plane (large yellow-filled circles). Different isochrones \citep{PCS04, PCS06} indicate the approximate loci of stars in the Galactic thin and thick disk, bulge, and halo.
}
\label{fig:model_grid}
\end{figure}

\section{Methodology\label{sect:method}}

\subsection{General considerations \label{sect:method-outline}}

To assess the influence of convection on the spectrophotometric properties of various types of stars, we computed SEDs, synthetic magnitudes, and colour indices using a set of 3D hydrodynamical and averaged $\xtmean{\mbox{3D}}$ model atmospheres that were calculated with the \COBOLD\ code (Sect.~\ref{sect:method-models-COBOLD}). For comparison, we utilized 1D hydrostatic models computed with the \LHD\ code (Sect.~\ref{sect:method-models-LHD}). Atmospheric parameters of the model atmospheres used were selected to span across those of stars located on the MS, main sequence turn-off (TO), subgiant branch (SGB), and RGB in various Galactic populations (Fig.~\ref{fig:model_grid}). These model atmospheres were used to compute wavelength-dependent surface fluxes, $F_{\lambda}$ (SEDs) for each given model atmosphere (Sect.~\ref{sect:method-seds+colors}). Finally, synthetic photometric magnitudes and colour indices were computed from the obtained SEDs in order to investigate the influence of convection on the spectrophotometric properties of these types of stars.

In this study we aimed at a strictly differential comparison of the observable properties predicted by the 3D hydrodynamical \COBOLD\ and 1D hydrostatic \LHD\ model atmospheres. Hence, the 3D hydrodynamical, averaged $\xtmean{\mbox{3D}}$, and 1D models used in our work were computed using identical atmospheric parameters (i.e. \Teff, $\log g$, and \moh), chemical composition, opacities, and equation of state.
To obtain SEDs that were subsequently  used for the computation of photometric magnitudes and colour indices, we used the code \NLTEIIID\ intended to solve the restricted non-local thermodynamic equilibrium (NLTE) problem in three spatial dimensions. Rather conveniently for our purposes, the code computes wavelength-dependent mean intensities and surface fluxes, $F_{\lambda}$ (in LTE, Sect.~\ref{sect:method-seds+colors}), which can be used to calculate photometric magnitudes and colour indices. We therefore tried to homogenize the input physics and numerical details of the model and spectral synthesis computations as much as possible so that the differences between the photometric magnitudes and colour indices could be traced back to the sole influence of convection. It is for these reasons that we refrained from referencing  one of the widely used 1D hydrostatic model atmospheres as 1D comparison models (such as, e.g. \ATLAS), because this choice would have led to additional differences in the assumptions and input physics between the 3D and 1D models, and thus  to additional systematic differences in the computed fluxes and photometric magnitudes and/or colour indices.

Despite this, one important remaining inconsistency is that both in the present paper and the companion Paper~I, opacities used in the computation of the model atmospheres and SEDs were slightly different (see Sect.~\ref{sect:method-seds+colors} for details). In addition, \COBOLD\ and \LHD\ model atmospheres, as well as their SEDs, were computed treating scattering as true absorption. Nevertheless, since we were interested in a differential 3D--1D comparison of the magnitudes and colour indices, the impact of these inconsistencies on our results is minor (see Sect.~\ref{sect:method-seds+colors} and Paper~I for further details and discussion).

\subsection{Model atmospheres\label{sect:method-models}}

\subsubsection{Three-dimensional hydrodynamical and averaged $\xtmean{\mbox{3D}}$ \COBOLD\ model atmospheres\label{sect:method-models-COBOLD}}

\begin{table}[tb]
        \caption{Three-dimensional hydrodynamical \COBOLD\ model atmospheres.\label{tab:3Dmodels}}
        \centering
        \setlength{\tabcolsep}{2pt}
        \begin{tabular}{lccccc}
                \hline
                Model   & \Teff & $\log g$ &    \moh\    &      Grid size , Mm       &      Grid resolution      \\
                &   K   &          &             &   $x \times y \times z$   &    $x\times y\times z$    \\
                \hline
                RGB \#1 & $4018\pm23$  & 1.5      & $\;\;\,0.0$ & 4610$\times$4610$\times$2780 & 140$\times$140$\times$150 \\
                RGB \#2 & $4490\pm10$  & 1.5      & $-2.0$      & 5390$\times$5390$\times$1840 & 160$\times$160$\times$200 \\
                RGB \#3 & $4477\pm11$  & 2.5      & $\;\;\,0.0$ & 573$\times$573$\times$243 & 160$\times$160$\times$200 \\
                RGB \#4 & $5024\pm12$  & 2.5      & $-2.0$      & 584$\times$584$\times$245 & 160$\times$160$\times$200 \\
                SGB \#1 & $4923\pm13$  & 3.5      & $\;\;\,0.0$ & 59.7$\times$59.7$\times$30.2 & 140$\times$140$\times$150 \\
                SGB \#2 & $5505\pm14$  & 3.5      & $-2.0$      & 49.0$\times$49.0$\times$35.9 & 140$\times$140$\times$150 \\
                TO \#1  & $5475\pm14$  & 4.0      & $\;\;\,0.0$ & 20.3$\times$20.3$\times$10.6    & 140$\times$140$\times$150 \\
                TO \#2  & $6534\pm13$  & 4.0      & $-2.0$      & 29.6$\times$29.6$\times$14.9    & 140$\times$140$\times$150 \\
                MS \#1  & $4982\pm13$  & 4.5      & $\;\;\,0.0$ & 4.94$\times$4.94$\times$2.48 & 140$\times$140$\times$141 \\
                MS \#2  & $6323\pm10$  & 4.5      & $-2.0$      & 7.00$\times$7.00$\times$4.02 & 140$\times$140$\times$150 \\
                \hline
        \end{tabular}
\end{table}

All  3D hydrodynamical stellar model atmospheres used in our study were computed as part of the CIFIST model atmosphere grid \citep{LCS09}. Simulations were performed using the \COBOLD\ code which solves time-dependent radiation-hydrodynamics equations in Cartesian geometry \citep[see][for details]{FSL12}. All model runs were long enough to cover between five and ten convective turnover timescales, as measured by the Brunt--Vais\"{a}l\"{a} timescale. Solar-scaled abundances from \citet{GS98} were used in the simulations, except that for CNO the values recommended by \citet{AGS05} were adopted: ${\rm A(C)}=8.41$, ${\rm A(N)}=7.80$, and ${\rm A(O)}=8.67$. Models at $\moh=-2.0$ were computed with a constant alpha-element enhancement of $\aoFe=+0.4$. Monochromatic \MARCS\ opacities \citep{GEK08} were divided into five and six opacity bins for the models at $\moh=0.0$ and $-2.0$, respectively. The scattering was treated as true absorption. All model simulations were done in LTE \citep[see][for more details on the model calculations]{LCS09}. A list of the 3D hydrodynamical models used in this work and their parameters are provided in Table~\ref{tab:3Dmodels}.

To speed up the computation of wavelength-dependent surface fluxes, for each 3D hydrodynamical model from Table~\ref{tab:3Dmodels} we selected a sample of 20 equidistantly spaced 3D model structures that were computed at different instants in time (snapshots). The snapshots were selected from the entire model simulation run such  that the most important statistical properties of the entire model sequence would be also preserved in the 20 snapshot subsample\footnote{The average effective temperature, \Teff, and its standard deviation, mean velocity at optical depth unity, mean velocity profile, and residual mass flux profile \citep[see e.g.][]{CL07,KSL13b}.}. Since each snapshot is characterized by slightly different total surface flux, and thus effective temperature, in Table~\ref{tab:3Dmodels} we also provide RMS variation in \Teff\ in each given snapshot ensemble.

We also used averaged $\xtmean{\mbox{3D}}$ model atmospheres which were computed by averaging individual 3D model structures in each of the 20 snapshot subsamples on surfaces of equal Rosseland optical depth. In fact, the averaged $\xtmean{\mbox{3D}}$ structures are 1D model atmospheres, thus they no longer carry the imprints of the horizontal fluctuations of hydrodynamical and thermodynamic quantities. Therefore, analysis of the 3D--$\xtmean{\mbox{3D}}$ differences helps us to assess the importance of horizontal fluctuations on the emergent surface fluxes, photometric magnitudes, and colour indices (see Sect.~\ref{sect:discuss-stars}).

\subsubsection{One-dimensional \LHD\ model atmospheres\label{sect:method-models-LHD}}

\begin{table}[tb]
        \caption{One-dimensional hydrostatic LHD\ model atmospheres.\label{tab:1Dmodels}}
        \centering
        \setlength{\tabcolsep}{2pt}
        \begin{tabular}{lccccc}
                \hline
                Model   & \Teff & $\log g$ &    \moh\     \\
                &   K   &          &              \\
                \hline
                RGB \#1 & 4020  & 1.5      & $\;\;\,0.0$  \\
                RGB \#2 & 4490  & 1.5      & $-2.0$       \\
                RGB \#3 & 4970  & 2.5      & $\;\;\,0.0$  \\
                RGB \#4 & 5020  & 2.5      & $-2.0$       \\
                SGB \#1 & 4920  & 3.5      & $\;\;\,0.0$  \\
                SGB \#2 & 5500  & 3.5      & $-2.0$       \\
                TO \#1  & 5930  & 4.0      & $\;\;\,0.0$  \\
                TO \#2  & 6530  & 4.0      & $-2.0$       \\
                MS \#1  & 4980  & 4.5      & $\;\;\,0.0$  \\
                MS \#2  & 6320  & 4.5      & $-2.0$       \\
                \hline
        \end{tabular}
\end{table}

One-dimensional hydrostatic model atmospheres were computed with the \LHD\ model atmosphere code \citep[][]{CLS08} in which convection was treated using the mixing-length theory in the formulation of \citet[][]{M78}. Stellar parameters, opacities, equation of state, and chemical composition used to compute the \LHD\ models were identical to those used in the \COBOLD\ simulations. In the standard set-up, the mixing-length parameter was set to $\mlp=1.0$. However, to test the sensitivity of the SEDs, photometric magnitudes, and colour indices on the choice of \mlp, additional \LHD\ models were computed with $\mlp=0.5$ and 2.0 (see Sect.~\ref{sect:discuss-other-effects}). We note, however, that the mixing length parameter was utilized only with the 1D \LHD\ hydrostatic models; in the 3D hydrodynamical (and averaged $\xtmean{\mbox{3D}}$) models the convective flux was modelled using the time-dependent equations of radiation hydrodynamics. A list of 1D \LHD\ model atmospheres and their parameters is given in Table~\ref{tab:1Dmodels}.

\subsubsection{Model atmospheres of the Sun and Vega\label{sect:method-models-Sun}}

In order to validate the accuracy and realism of our model atmospheres, we compared synthetic solar SEDs computed using 3D hydrodynamical, averaged $\xtmean{\mbox{3D}}$, and 1D hydrostatic model atmospheres with the observed solar flux spectrum. For this, we used the \COBOLD\ and \LHD\ models of the Sun from \citet{SPC15}. These are currently the most advanced solar models computed using the \COBOLD\ and \LHD\ codes \citep[see][for details]{SPC15}. Therefore, by utilizing them we tried to ensure that the best theoretical models were used to compare their predictions with the observations. 

Similar comparison was also made in the case of Vega. Since its atmosphere is fully radiative, it was neither possible nor desirable to compute a 3D hydrodynamical model of Vega because the flux of such a model should be identical to that of the 1D \LHD\ model atmosphere. Thus, to compare the observed and theoretical SEDs, we used the 1D~\LHD\ and \ATLAS\ models of Vega (the former model was also used to compute the magnitude zero points for Vega, see Sect.~\ref{sect:method-seds+colors}). The atmospheric parameters of Vega adopted for the 1D~\LHD\ model atmosphere calculations were $\Teff=9550$~K, $\log g=3.95$, and $\moh=-0.5$, with an enhancement in $\alpha$-element abundances of $\aoFe=+0.2$. Just as for the cooler \COBOLD\ and \LHD\ models in our sudy, the elemental abundances used in the computation of the Vega 1D~\LHD\ model were taken from \citet{GS98}, with the updates on CNO abundances from \citet{AGS05}. The equation of state used to produce a 1D \LHD\ model of Vega was identical to that utilized in the computation of all other \COBOLD\ and \LHD\ models listed in Tables~\ref{tab:3Dmodels} and \ref{tab:1Dmodels}. We used a 12-bin opacity table based on Hayek's approximation in which scattering was treated as true absorption in the deeper layers and  was ignored in the outer ones (see Paper~I for further discussion on scattering-related issues). The \ATLAS\ SED of Vega was taken from \citet{CK94}.

\subsection{Spectral energy distributions, photometric magnitudes, and colour indices\label{sect:method-seds+colors}}

The wavelength-dependent surface fluxes, $F_{\lambda}$, of the 3D hydrodynamical, averaged $\xtmean{\mbox{3D}}$, and 1D hydrostatic model atmospheres were computed using \NLTEIIID\ package. This package is designed to compute NLTE departure coefficients, $b_{\rm i}(x,y,z)$, at each geometrical position $(x,y,z)$ in the 3D model atmosphere box and for each energy level $i$ of a given model atom \citep[for details regarding the \NLTEIIID\ code see][]{SPC15}. The \NLTEIIID\ code also computes wavelength-dependent mean intensity at each geometrical position in the model atmosphere, $J_{\lambda}(x,y,z)$, as well as the emergent flux at the top of the atmosphere, $F_{\lambda}$, which can be used to calculate photometric magnitudes and colour indices. Continuum opacities used in the \NLTEIIID\ code were computed with the {\tt IONDIS} and {\tt OPALAM} routines, which are also employed in the \Linfor\ spectral synthesis code \citep{Linfor2015}. These are different from the \MARCS\ continuum opacities that were used to produce the binned opacity tables employed in the \COBOLD\ and \LHD\ simulations. Line opacities used in the computation of SEDs were taken into account by utilizing {\tt LITTLE} opacity distribution functions (ODFs) from the \ATLAS\ model atmosphere package \citep{CK03}. Both choices lead to inconsistencies between the opacities used in the calculations of SEDs (and thus  photometric magnitudes and colour indices) and those employed in the computation of the \COBOLD\ and \LHD\ models.

As a consequence, wavelength-integrated surface fluxes obtained from the SEDs that were computed using different model atmospheres were slightly inconsistent, even though the effective temperatures of all models were identical. To correct for this mismatch, we re-normalized the SEDs of the 3D, $\xtmean{\mbox{3D}}$, and 1D models to their nominal effective temperatures using wavelength-independent scaling factors. This was not done in \citet{BCL17} and Paper~I. The differences in the 3D--1D corrections due to the different approaches used in the two studies were in fact very small and did not exceed 0.001\,mag.

For the calculation of magnitudes and colour indices we used standard \ATLAS\ {\tt LITTLE} ODFs computed with the microturbulence velocity set to $\xi_{\rm mic}=2$~km/s. However, in order to assess a possible influence of this choice on the 3D--1D corrections for magnitudes and colour indices, we also used ODFs computed with $\xi_{\rm mic}=1.0$~km/s (Sect.~\ref{sect:discuss-other-effects}). The wavelength-dependent surface fluxes, $F_{\lambda}$ (SEDs), were calculated at 1048 wavelength points spread non-equidistantly between 133.5 and 99\,799~nm. The blue wavelength limit was chosen to allow for the computation of near-UV photometric magnitudes and colour indices (see Paper~I).

From the obtained SEDs we computed photometric magnitudes and colour indices in the Johnson-Cousins $UBVRI$, 2MASS $JHK_{\rm s}$, and Str\"{o}mgren $uvby$ photometric systems. The $UBVRIJHK_{\rm s}$ system was chosen because it is well defined and its bandpasses are located in the wavelength range that extends over those of other widely used broad-band photometric systems, such as SDSS $ugriz$. Therefore, by estimating the influence of convection on the $UBVRIJHK_{\rm s}$ magnitudes and colour indices one can obtain a rough idea regarding the size of 3D--1D corrections  expected at similar wavelengths for other broad-band photometric systems. The medium-band Str\"{o}mgren $uvby$ photometric system was chosen because its bandpasses are mostly located in the blue part of the spectrum where, as we will see in Sect.~\ref{sect:discuss-stars-seds}, the influence of 3D effects is greatest. On the other hand, its filter transmission curves are  relatively narrow.  At certain wavelengths where the 3D--1D differences are largest, the Str\"{o}mgren's magnitudes and colour indices may be affected more strongly than those in the $UBVRIJHK_{\rm s}$ system,  thereby allowing us to assess the influence of convection on magnitudes and colour indices in the photometric bands of different width.

Synthetic magnitudes computed in both photometric systems were normalized with respect to Vega. For this, we followed \citet{BM12} and assumed the $UBVRI$ magnitudes of Vega equal to 0.03, while for the 2MASS bands we used $J=0.001$, $H=-0.005$, and $K_{\rm s}=0.001$, in order to be consistent with the zero-magnitude fluxes provided by \citet{CWM03}. For the Str\"{o}mgren bands, we utilized Vega magnitudes from \citet{MA07}. The filter transmission curves for the computation of the zero points and photometric magnitudes were taken from \citet{BM12} for the Johnson-Cousins $UBVRI$ bands and from \citet{CWM03} for the 2MASS $JHK_{\rm s}$ bands. For the Str\"{o}mgren $uvby$ system, we used the filter transmission curves from \citet{MA06}. 

At variance with \citet{BCL17} and Paper~I, where we assumed an angular diameter for each model atmosphere and applied to the synthetic flux the same zero points that were applied to colour indices, in this paper we have chosen to determine the zero points of the synthetic colour indices using our 1D~\LHD\ model atmosphere of Vega. The magnitude zero points in the different filter bands were obtained using the SED computed with the 1D~\LHD\ model atmosphere of Vega and applied to synthetic magnitudes calculated with the 3D hydrodynamical, averaged $\xtmean{\mbox{3D}}$, and 1D hydrostatic models. The obtained magnitude zero points were in good agreement with those determined in other studies; for example,  the differences between our zero points and those determined by \citet{CV14} did not exceed 0.01~mag. 

We note, however, that in both papers we were interested in a strictly differential comparison of fluxes, photometric magnitudes, and/or colour indices, i.e.  the differences between these quantities computed using 3D and 1D model atmospheres. In this situation, the choice of particular zero points is irrelevant because in the differential comparison of 3D--1D magnitude differences they cancel out. Considering the limitations in the model physics due to our treatment of scattering (see Paper~I for details), we discourage the use of  our colour indices or magnitudes and recommend using only the 3D--1D colour corrections.

\section{Results and discussion\label{sect:discussion}}

\subsection{Spectral energy distribution of the Sun\label{sect:discuss-sun}}

\begin{figure}[tb]
        \centering
        \includegraphics[width=\columnwidth]{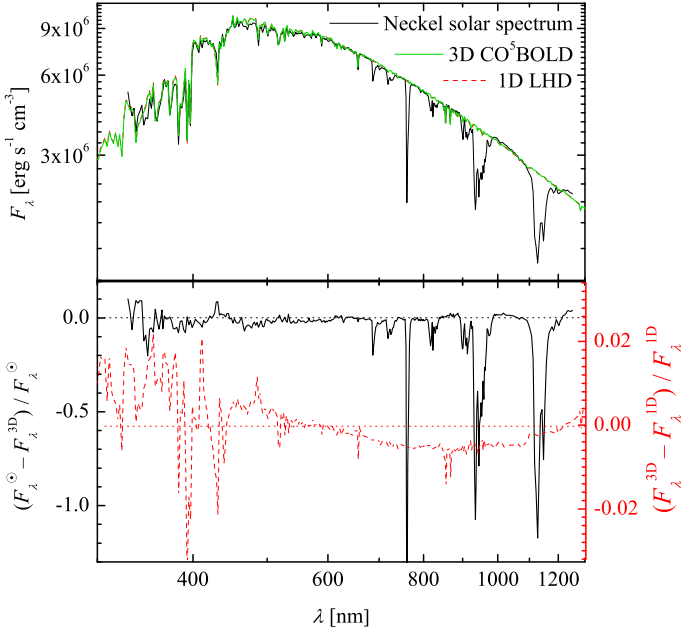}
        \caption
        {Comparison of theoretical and observed SEDs of the Sun. \textbf{Top:} Solar SEDs computed using 3D hydrodynamical \COBOLD\ (green solid line) and 1D \LHD\ (red dotted line) model atmospheres, plotted together with the observed absolute solar flux spectrum from \citet[][black solid line]{N99}. \textbf{Bottom}: Relative differences between the SEDs: observed--3D  (black solid line) and 3D--1D  (red dashed line, scale on the right $y$-axis).}
        \label{fig:SEDs_Sun}
\end{figure}

\begin{figure}[tb]
        \centering
        \includegraphics[width=\columnwidth]{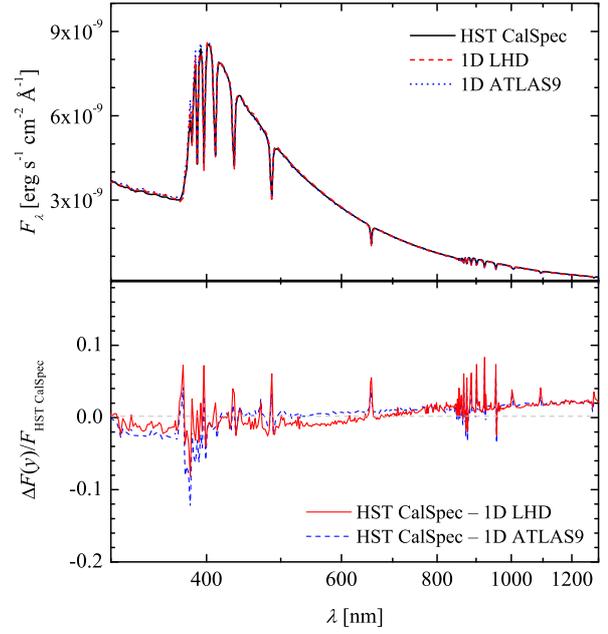}
        \caption
        {Comparison of theoretical and observed SEDs of Vega. \textbf{Top:} SEDs of Vega computed using 1D hydrostatic \LHD\ (red dotted line) and \ATLAS\ (blue dashed line) model atmospheres, plotted together with the \calspec\ absolute calibrated flux spectrum of Vega from the \calspec\ library (black solid line). The  close overlap of the three lines makes it difficult to distinguish between them. \textbf{Bottom}: Relative differences between the SEDs:  \calspec\ -- \LHD\ (red dotted line) and \calspec\ -- \ATLAS\ (blue dashed line).}
        \label{fig:SEDs_Vega}
\end{figure}

A comparison between the synthetic solar SEDs computed using 3D hydrodynamical \COBOLD\ and 1D hydrostatic \LHD\ model atmospheres with the observed solar absolute calibrated flux spectrum of \citet{NL84} and taken from the {\tt ftp} site as provided in \citet[][]{N99} is shown in Fig.~\ref{fig:SEDs_Sun}. The observed spectrum was re-sampled to the resolution of the synthetic SEDs, while the latter were scaled to the distance of one astronomical unit and one solar radius. In the case  of the 3D hydrodynamical and averaged $\xtmean{\mbox{3D}}$ models, the SEDs were computed using the 20-snapshot ensemble, as described in Sect.~\ref{sect:method-models-COBOLD} \citep[see also][]{SPC15}. We emphasize that this is a comparison on an absolute scale, i.e. no fitting was performed. 

As seen in Fig.~\ref{fig:SEDs_Sun}, the differences between SEDs computed using 3D hydrodynamical and 1D hydrostatic model atmospheres are generally very small and never exceed a few percent. Nevertheless, certain systematic differences are apparent as well. For example, a 3D hydrodynamical model produces more flux in the blue and less flux in the red part of the spectrum (we note that the synthetic SED of the 1D model was normalized to produce the same wavelength-integrated flux as that of the 3D hydrodynamical model). The observed SED was not corrected for telluric absorption, thus larger deviations between the observed and synthetic SEDs are apparent in the red and infrared parts of the spectrum. Towards the blue, a slight tendency towards excessive flux in the synthetic SEDs is apparent, with the exception of a region around the H$\gamma$ line. Nevertheless, the correspondence between calculated and observed solar SEDs is in general very satisfactory, in particular considering the rather sensitive ultraviolet wavelength region.

\subsection{Spectral energy distribution of Vega\label{sect:discuss-vega}}

A comparison of the observed and synthetic SEDs of Vega is shown in Fig.~\ref{fig:SEDs_Vega}. For this, we used 1D hydrostatic model atmospheres computed with the \ATLAS\ and \LHD\ codes. Because of inconsistencies in the opacities used to compute the 1D \LHD\ model of Vega and its SED (Sect.~\ref{sect:method-seds+colors}), the \LHD\ SED of Vega was renormalized to produce the same flux as that of the \ATLAS\ model. The observed SED of Vega was taken from the \calspec\  library\footnote{http://www.stsci.edu/hst/observatory/cdbs/calspec.html} and was re-sampled to the spectral resolution of synthetic SEDs. For the comparison, synthetic SEDs were scaled to the distance of Vega \citep[7.69\,pc,][]{VL07} and the radius of Vega \citep[2.818~R$_{\odot}$,][]{YPK10}. The scaled SEDs were then directly compared with the \calspec\ SED of Vega, i.e. no fitting was done.

As in the case of the Sun, the agreement between the observed and synthetic SEDs is very good in the entire wavelength range covered in Fig.~\ref{fig:SEDs_Vega}. Larger differences are seen in the regions around strong lines (mostly those of hydrogen), which in extreme cases may reach  $10$\,\%. Some systematics are seen, too: the flux in the \calspec\ SED is slightly lower in the UV and somewhat higher in the IR parts of the spectrum. Nevertheless, the typical differences are small and do not exceed $\pm 2$~\%. The fluxes of the two 1D hydrostatic models, \ATLAS\ and \LHD, are also very similar although there are small differences, on the level of a few percent, in the UV and optical parts of the spectrum.

As a final remark, it is worth mentioning that Vega is too complex an object to model with the standard model atmospheres. For example, the above-mentioned IR excess of the observed spectrum over that of the LHD model is expected because Vega is known to possess a circumstellar disk (see discussion in \citealt{B14}). In addition, Vega is a rapid rotator and this may also have an impact on the structure of its atmosphere and, thus, on the observable properties. Finally, because of its rapid rotation there may be other effects, for example, those related to gravity darkening that are not yet taken into account in the current model atmospheres of Vega. Therefore, certain systematic differences between the observed \calspec\ spectrum and the synthetic \LHD\ and \ATLAS\ SEDs of Vega are to be expected. What our comparison shows is that these differences are nevertheless relatively small.

\begin{figure}[tb]
        \centering
        \includegraphics[width=\columnwidth]{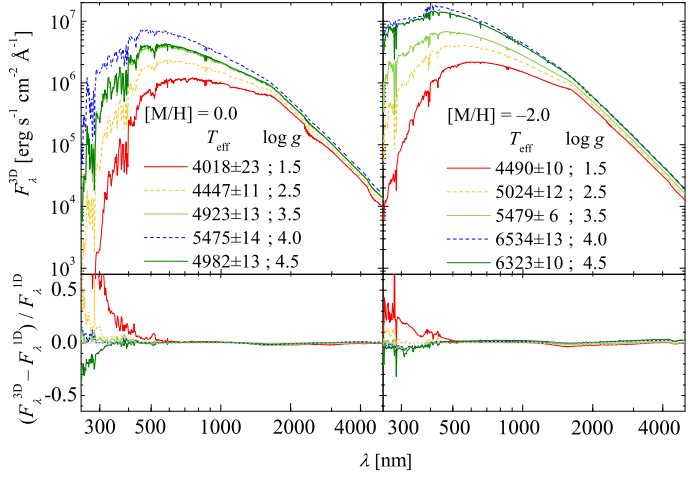}
        \caption
        {\textbf{Top:} SEDs corresponding to the 3D hydrodynamical \COBOLD\ models listed in Table~\ref{tab:3Dmodels}. \textbf{Bottom:} Relative flux difference between the SEDs computed using 3D hydrodynamical \COBOLD\ and 1D hydrostatic \LHD\ model atmospheres. Left panels: $\moh=0.0$; right panels: $\moh=-2.0$.}
        \label{fig:SEDs_stars}
\end{figure}

\begin{figure}[tb]
        \centering
        \includegraphics[width=\columnwidth]{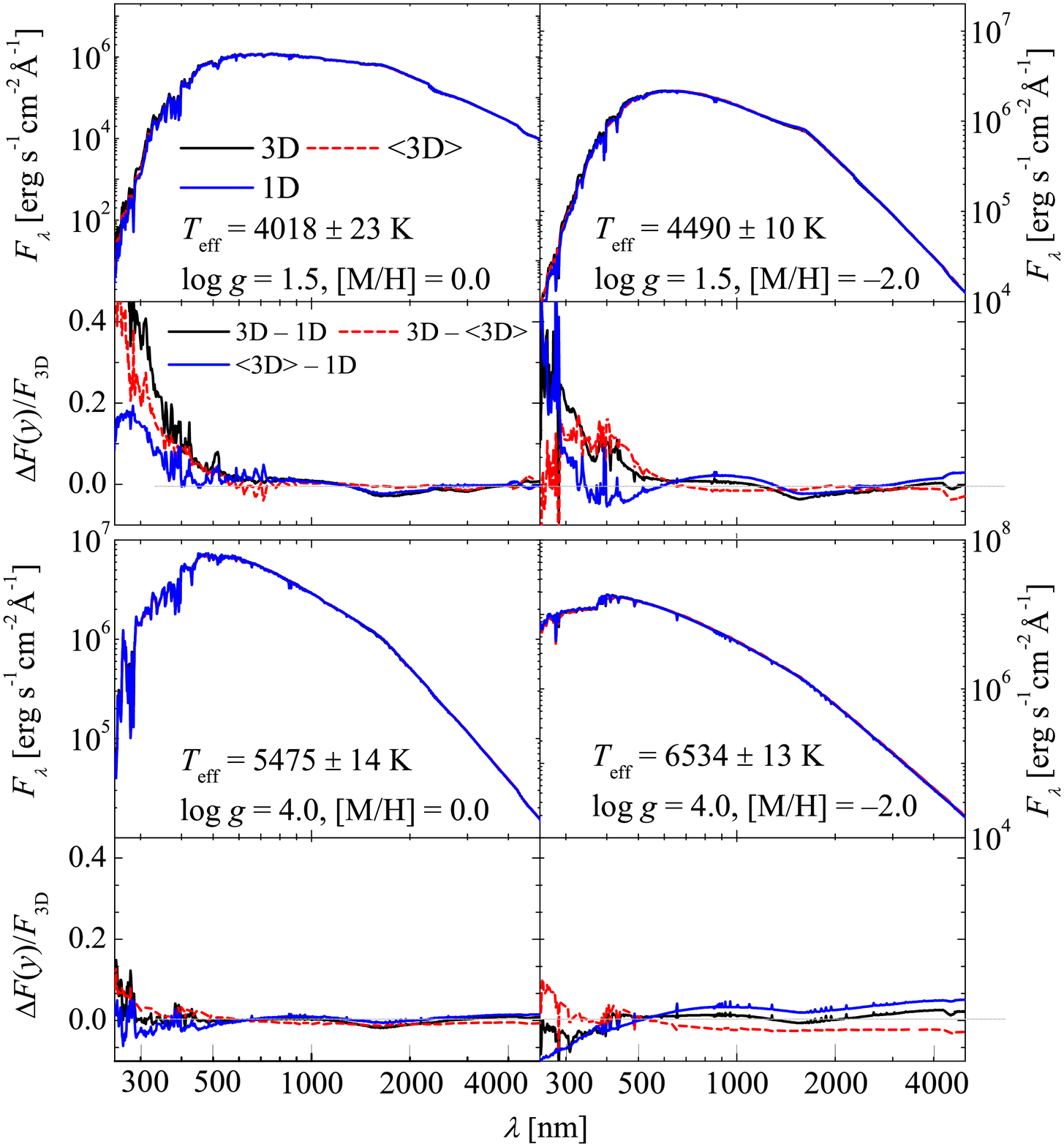}
        \caption
        {SEDs of the RGB (top row) and TO (bottom row) models computed using 3D hydrodynamical, averaged $\xtmean{\mbox{3D}}$, and 1D hydrostatic model atmospheres, at $\moh=0.0$ (left) and $\moh=-2.0$ (right). \textbf{Top panels:} SEDs corresponding to different models. \textbf{Bottom panels:} Relative differences between the SEDs shown in the corresponding top panels. All SEDs were re-normalized to yield the same effective temperature.}
        \label{fig:SEDs_stars_selected}
\end{figure}

\begin{figure}[tb]
        \centering
        \includegraphics[width=\columnwidth]{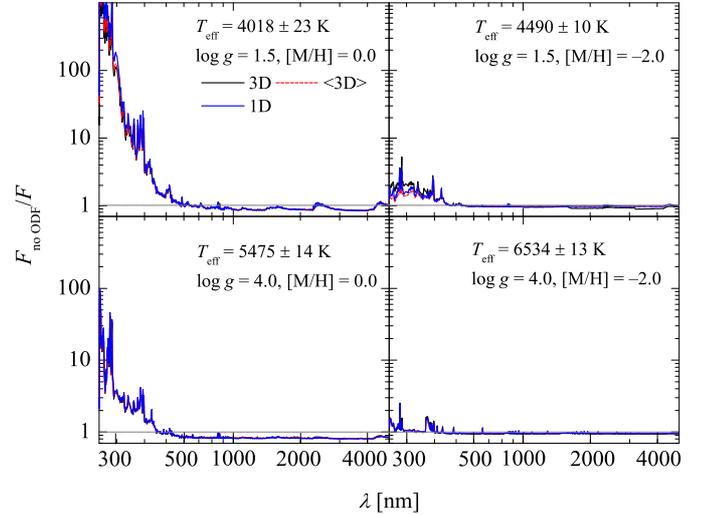}
        \caption
        {Relative differences between the SEDs of RGB (top row) and TO (bottom row) stars computed with and without ODFs using 3D hydrodynamical, averaged $\xtmean{\mbox{3D}}$, and 1D hydrostatic model atmospheres, at $\moh=0.0$ (left) and $\moh=-2.0$ (right).}
        \label{fig:SEDs_diff_ODF-noODF}
\end{figure}

\subsection{Spectral energy distributions and photometric properties of MS, TO, SGB, and RGB stars\label{sect:discuss-stars}}

\subsubsection{Spectral energy distributions\label{sect:discuss-stars-seds}}

Spectral energy distributions computed with the 3D hydrodynamical model atmospheres, as well as differences between the SEDs obtained using 3D hydrodynanical and 1D hydrostatic models, are shown in Fig.~\ref{fig:SEDs_stars}. A common tendency is that 3D hydrodynamical models tend to predict more flux in the UV and less in the IR part of the spectrum. In the UV, the 3D--1D differences quickly diminish with increasing effective temperature, i.e. they are largest in cool MS and RGB stars, and smallest in TO stars. The influence of gravity, on the other hand, is small.

\begin{figure*}[tb]
        \centering
        \includegraphics[width=16.5cm]{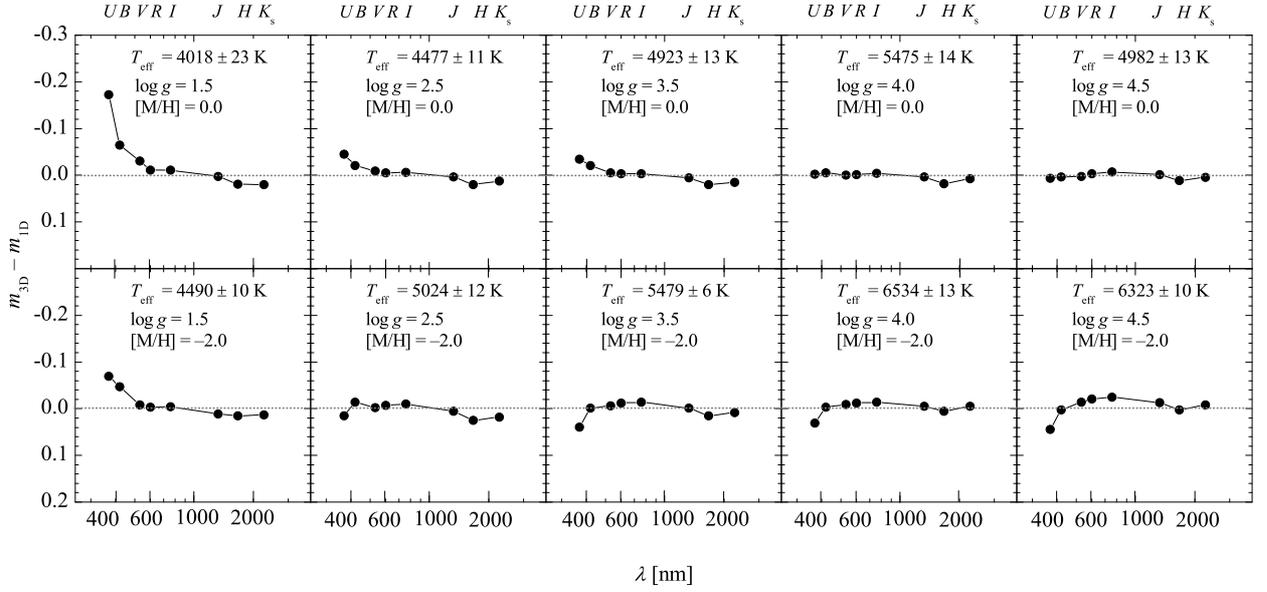}
        \caption
        {The 3D--1D magnitude differences in the $UBVRIJHK_{\rm s}$ photometric system computed using 3D hydrodynamical \COBOLD\ and 1D hydrostatic \LHD\ models, at $\moh=0.0$ and --2.0.}
        \label{fig:col-dif_JCG}
\end{figure*}

\begin{figure*}[tb]
        \centering
        \includegraphics[width=13.8cm]{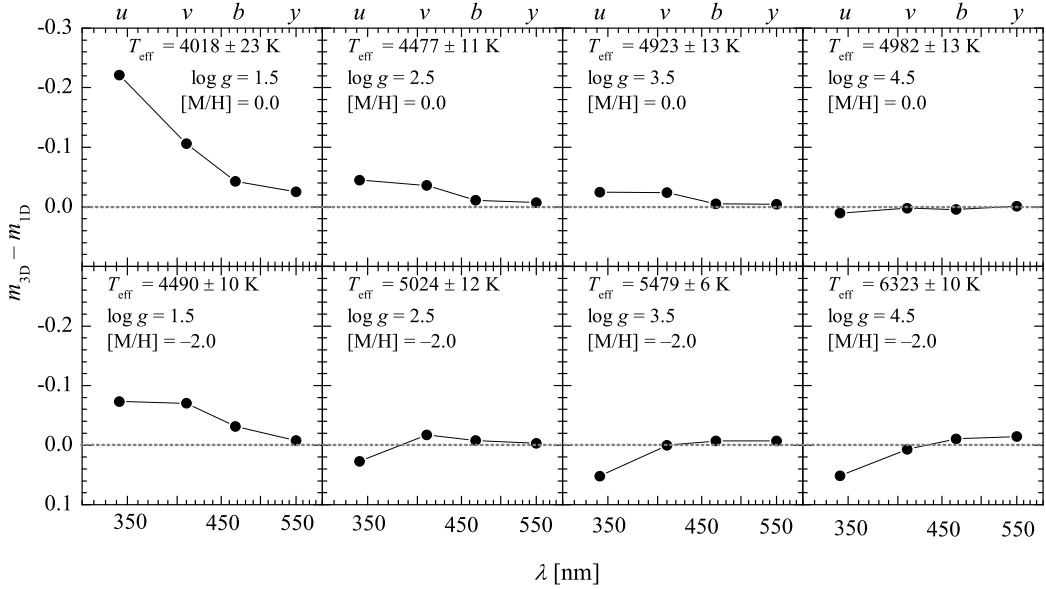}
        \caption
        {Same as  Fig.~\ref{fig:col-dif_JCG}, but in the Str{\o}mgren photometric system.}
        \label{fig:col-dif_Str}
\end{figure*}

Following our considerations presented in Appendix~\ref{sect_app:color-corr-model}, higher flux in the UV and lower in the IR should be expected in the presence of horizontal temperature fluctuations alone, i.e. even if their effect on the opacities is not taken into account. This is confirmed by a comparison of realistic SEDs computed using the 3D, averaged $\xtmean{\mbox{3D}}$, and 1D model atmospheres of RGB and TO stars (Fig.~\ref{fig:SEDs_stars_selected}). In the coolest red giants, higher flux in the UV is mostly caused by horizontal temperature fluctuations in the 3D hydrodynamical model atmosphere, as traced by the 3D--$\xtmean{\mbox{3D}}$ differences in Fig.~\ref{fig:SEDs_stars_selected} (top row). In the IR, the horizontal fluctuations and differences between the average temperature profiles in the $\xtmean{\mbox{3D}}$ and 1D model atmospheres play a comparable role, with the joint effect typically leading to lower flux in the IR ($\lambda \geq 1000$\,nm). Similar behaviour is seen in the SEDs of MS stars too, although the total effect, especially in the UV, is smaller (Fig.~\ref{fig:SEDs_stars_selected}, bottom row). In part, this is caused by the negative $\xtmean{\mbox{3D}}$--1D differences, especially in the low-metallicity model. Our tests have shown that continuum flux in the UV typically forms in the deeper atmospheric layers where horizontal temperature fluctuations are larger, while the formation of IR continuum is confined to the outer layers where fluctuations are smaller. In addition, differences between temperature profiles of the averaged $\xtmean{\mbox{3D}}$ and 1D models are also larger in the deeper atmosphere. All together, this leads to a larger cumulative effect in the UV than in the IR and, consequently, to larger 3D--1D differences in the UV. 

As far as the role of opacities is concerned, the largest contribution comes from the bound-bound transitions, which in the computation of SEDs were taken into account using \ATLAS\ ODFs. This is illustrated in Fig.~\ref{fig:SEDs_diff_ODF-noODF}, which displays differences between the SEDs computed using 3D, $\xtmean{\mbox{3D}}$, and 1D models, with the ODFs switched on and off. The results of this test show that line opacities (often referred to as `line blanketing') indeed have a huge influence on the resulting SEDs of solar metallicity models. Their contribution gradually becomes dominant towards the shorter wavelengths ($\lambda\leq 600$\,nm) where spectral lines of various metals become more abundant. As expected, at $\moh=-2.0$ the effect of line opacity is smaller although still non-negligible. We recall, however, that this is the situation when scattering is treated as true absorption; the overall picture may be different when scattering is properly taken into account.

\begin{figure}[tb]
        \centering
        \includegraphics[width=8.2cm]{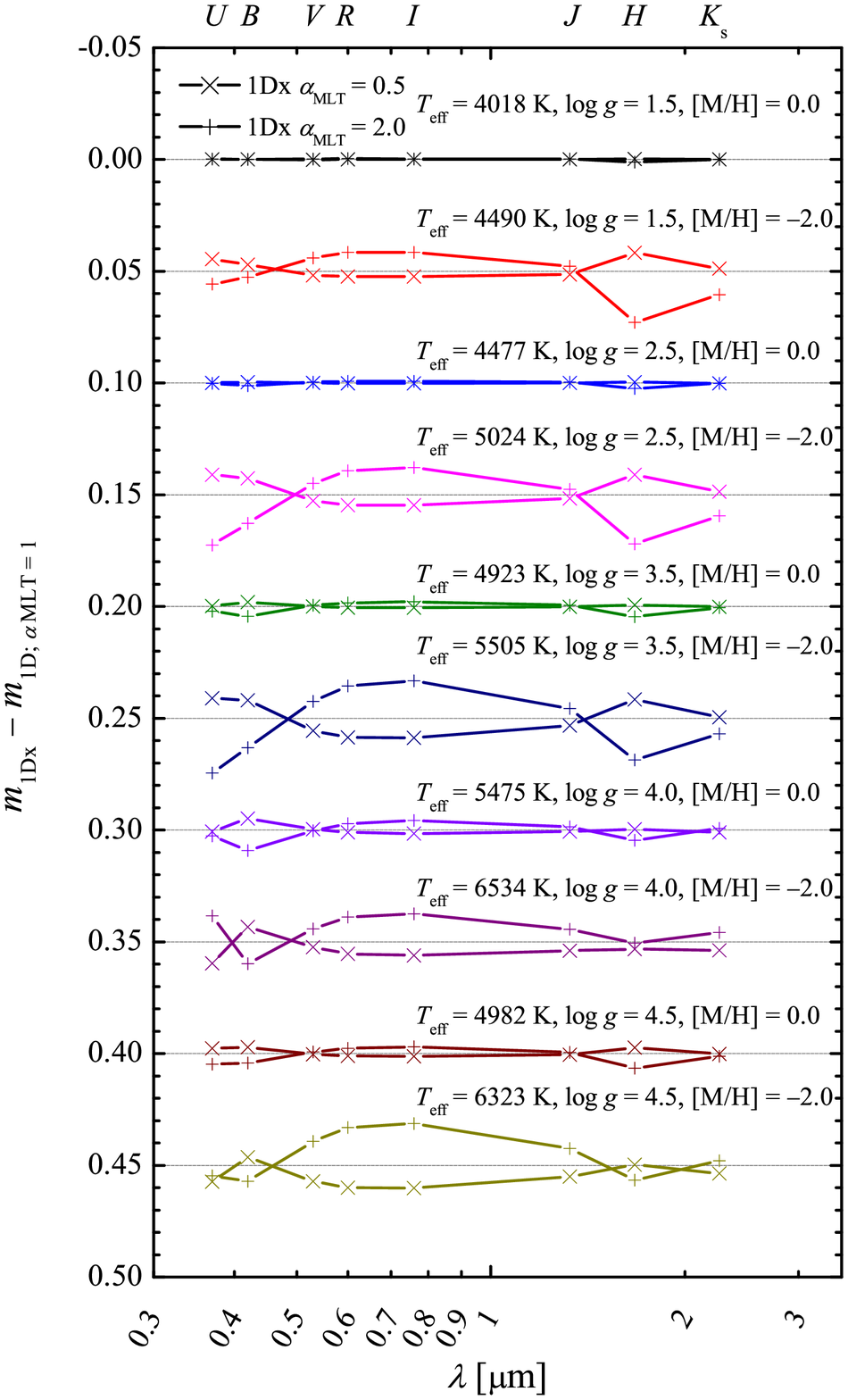}
        \caption
        {Differences in the $UBVRIJHK_{\rm s}$ magnitudes obtained from the 1D~\LHD\ models that were computed using different mixing length parameters, plotted as differences with respect to the models computed with $\mlp=1.0$. Dotted horizontal lines mark zero magnitude differences. The $y$-scale increases downwards and the zero difference lines are shifted incrementally by +0.05\,mag from top to bottom.}
        \label{fig:amlt-mag}
\end{figure}

\begin{figure}[tb]
        \centering
        \includegraphics[width=8.2cm]{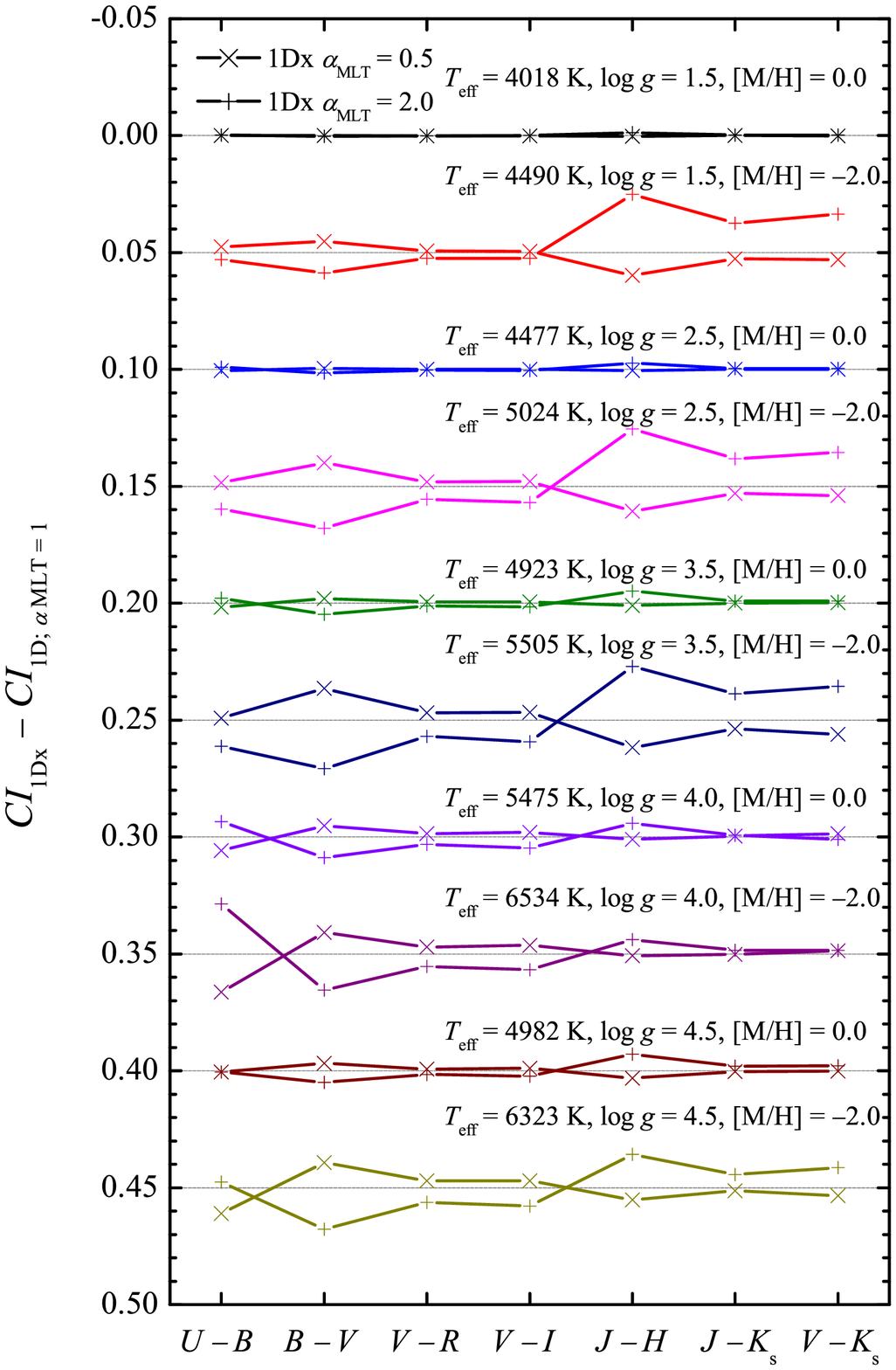}
        \caption
        {Same as  Fig.~\ref{fig:amlt-mag}, but for the colour indices.}
        \label{fig:amlt-col}
\end{figure}

\subsubsection{Photometric magnitudes and colour indices\label{sect:discuss-stars-colors}}

A glance at the differences between $UBVRIJHK_{\rm s}$ magnitudes computed using the 3D and 1D model atmospheres (Fig.~\ref{fig:col-dif_JCG}) reveals similar patterns to those seen in the comparison of the SEDs (Sect.~\ref{sect:discuss-stars-seds}). In particular, (a) 3D models tend to produce more flux in the UV and less in the IR; (b) differences in the magnitudes and colour indices, especially those with filter bands located in the blue-UV part of the spectrum, become smaller with increasing effective temperature of the model atmosphere; (c) at the highest effective temperatures, the differences are somewhat more pronounced in the low-metallicity models. Interestingly, in the metal-poor models the initially positive 3D--1D differences in the bluest bands quickly decrease in their absolute size with increasing effective temperature and gradually become negative for models characterized with the highest effective temperatures. A similar pattern is  seen in the case of magnitudes and colour indices in the Str\o{}mgren system (Fig.~\ref{fig:col-dif_Str}). Overall, the 3D--1D differences in the two photometric systems are small and generally are well within $\pm 0.1$\,mag,  for magnitudes and for colour indices. As discussed in Sect.~\ref{sect:discuss-stars-seds}, the 3D--1D differences in the magnitudes and colour indices are mostly caused by horizontal temperature fluctuations in the 3D hydrodynamical model atmospheres which affect both the source function and opacities.

\subsubsection{The impact of various modelling assumptions on synthetic colours\label{sect:discuss-other-effects}}

\paragraph{The mixing length parameter.}

It is known that the choice of the mixing length parameter, \mlp, which controls the efficiency of convective mixing and has a direct influence on the thermal structure of the 1D hydrostatic model atmosphere, may affect the observable properties of a given model atmosphere. In the context of our study, \mlp\ was used in the computations of 1D \LHD\ model atmospheres which were utilized in the differential 3D--1D analysis of photometric magnitudes and colour indices. Our standard \LHD\ models were computed with $\mlp=1.0$. To investigate the influence of this choice on the spectrophotometric properties of the \LHD\ models, we computed a set of additional \LHD\ models with $\mlp=0.5$ and 2.0. In Figs.~\ref{fig:amlt-mag} and \ref{fig:amlt-col} we show differences in the photometric magnitudes and colour indices of the \LHD\ models computed with $\mlp=0.5$ and 2.0 and those computed with $\mlp=1.0$. The resulting differences in the 3D--1D corrections will be of the same size.

As expected, in the metal-poor models different choices of \mlp\ lead to larger differences in the photometric magnitudes and colour indices because in this case the convective envelope penetrates closer to the optical surface than it does in the solar metallicity models \citep[see e.g.][]{KLS13a}. Despite this, the differences are small, and for both magnitudes and colour indices are confined to $\approx \pm0.025$\,mag.

\paragraph{The role of scattering.}

The \COBOLD\ and \LHD\ model atmospheres used in our study were computed by treating scattering as true absorption \citep{LCS09,FSL12}. The identical procedure was  applied when computing SEDs with the \NLTEIIID\ code from the 3D, averaged $\xtmean{\mbox{3D}}$, and 1D models. The main reasons behind this choice was that it reduces the complexity of radiative transfer (RT) calculations and increases the computational speed dramatically\footnote{While the \NLTEIIID\ code allows the SEDs to be computed using the full treatment of scattering, all \COBOLD\ models in the CIFIST grid were computed by treating scattering as true absorption (see Paper~I for a further discussion on the possible implications related to this choice).}. Several recent studies have  focused on the validity of this approach in the context of 3D hydrodynamical modelling of stellar atmospheres with different model atmosphere codes \citep{HAC10,CHA11,LS12}. While the obtained results are somewhat inconclusive regarding how much the thermal structure of the atmosphere may be affected by this choice, it is clear that scattering makes an important contribution to the emergent radiation field in the blue-UV part of the spectrum, especially in the metal-poor regime. 

In our companion paper we therefore investigated a possible impact of this simplified approach on the spectrophotometric properties of the model atmospheres computed with the \COBOLD\ code (Paper~I). For this, we used two \COBOLD\ models computed with identical atmospheric parameters but different treatment of scattering, (a) with scattering treated as true absorption, and (b) with scattering treated in the Hayek approximation, which is deemed to be a sufficiently accurate substitute for a more complex approach where scattering is treated fully consistently. These models were used to compute photometric colour indices, again with (a) scattering treated as true absorption, and (b) by performing fully consistent computations of scattering. We refer the reader to Paper~I for the in-depth discussion of the results obtained in this test, and here we  only briefly note that scattering was indeed found to be important in the blue-UV part of the spectrum. Fortunately, the 3D--1D differences in the colour indices computed using different approaches in the treatment of scattering were always consistent to within $0.03$\,mag, even when the bluest colour indices were involved (see Sect.~2.3 in Paper~I).

\begin{figure}[tb]
        \centering
        \includegraphics[width=\columnwidth]{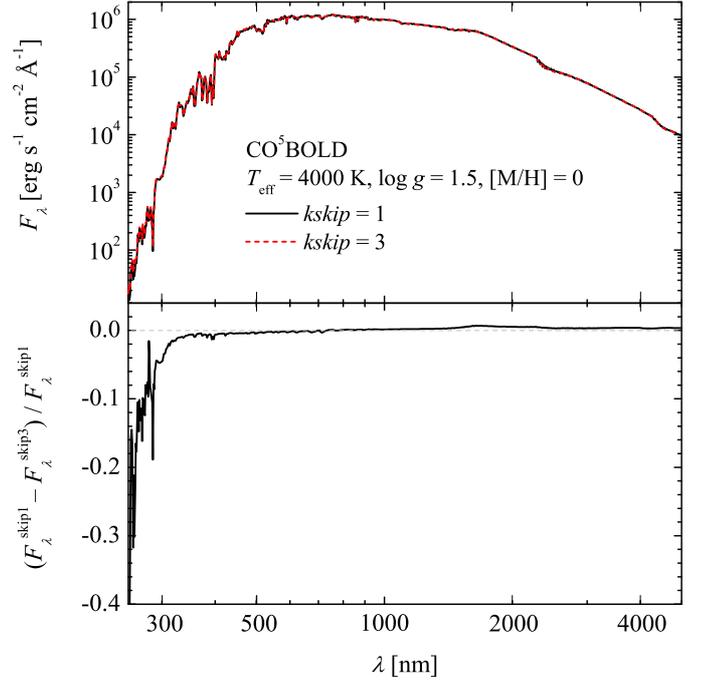}
        \caption
        {\textbf{Top:} SEDs computed using the original grid of the model atmosphere ($kskip = 1$) and a coarser grid covering each third cell of the original grid ($kskip=3$). A single snapshot of the 3D hydrodynamical model of a red giant was used in the computation of both SEDs. \textbf{Bottom:} Differences between the two SEDs shown in the top panel.}
        \label{fig:skip-par}
\end{figure}

\paragraph{Radiative transfer computations.}

To speed up the RT computations, the radiative field in the 3D hydrodynamical models is frequently computed on a coarser grid than that used in the model simulations, for example, at each third horizontal point of the original grid \citep[e.g.][]{KSL13b, DKS13}. This effectively reduces the number of grid points where the radiation field is computed significantly, which speeds up the computations. Normally, this procedure does not introduce significant differences even in the computation of spectral line profiles, which are typically rather sensitive to the properties of the grid resolution \citep{KSL13b}.

In the present study we did not use this simplification. Instead, the emergent flux was computed at each grid point of the model atmosphere. Nevertheless, we performed RT calculations with a coarser grid using only each third horizontal point of the original grid to estimate the possible effect of this frequently used choice on the emergent flux. For this test, we used a single snapshot of the \COBOLD\ model of a red giant star with $\Teff = 4000$\,K, $\log g = 1.5$, and $\moh = 0.0$. This choice was motivated by the fact that in these cool and low-gravity red giants horizontal temperature fluctuations are characterized by very steep gradients and large amplitudes, which makes these stars good targets for such a test. 

The SEDs computed using the original and coarser numerical grids are shown in Fig.~\ref{fig:skip-par}. The differences typically do not exceed 0.5\% in the optical-IR range but they increase rapidly with decreasing wavelength at $\sim300$\,nm and may grow alarmingly large below $\sim250$\,nm. This, however, has a negligible effect on the magnitudes and colour indices in the $UBVRIJHK_{\rm s}$ and Str{\o}mgren systems as their filter transmission curves are located at wavelengths longer than $300$\,nm.

\paragraph{Effects related to opacities.}

Two aspects should be mentioned here. First, as  was already discussed in Sect.~\ref{sect:method-seds+colors}, opacities used in the computation of the model atmospheres and those utilized in the calculation of SEDs were slightly inconsistent: the \COBOLD\ and \LHD\ models were computed using binned \MARCS\ opacities, while the SEDs and magnitudes and colour indices were computed using monochromatic continuum opacities from the Linfor3D codes together with the line opacities using ODFs from the \ATLAS\ package. Fortunately, in the context of differential comparison of the 3D and 1D model predictions these inconsistencies should not pose a serious problem since their effects will be expected to largely cancel out.

\begin{figure}[tb]
        \centering
        \includegraphics[width=\columnwidth]{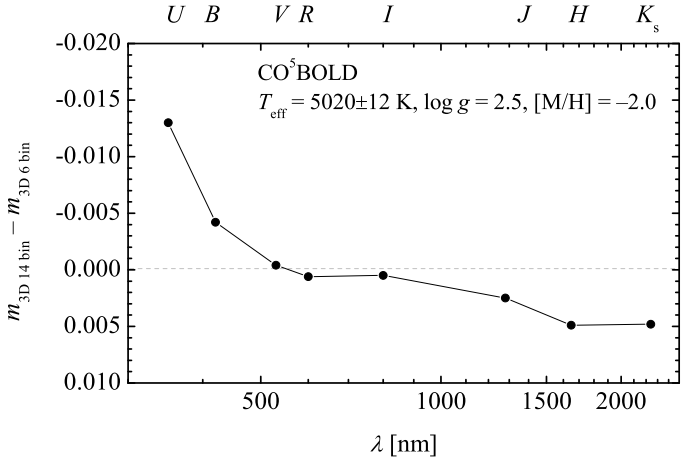}
        \caption
        {Effects of opacity binning on the photometric magnitudes in the $UBVRIJHK_{\rm s}$ system. The plot shows differences in the magnitudes of two 3D hydrodynamical \COBOLD\ models computed using opacities grouped into 14 and 6 opacity bins.}
        \label{fig:opacity-binning}
\end{figure}

Second, we used binned opacities in the computation of \COBOLD\ and \LHD\ model atmospheres. The main advantage of using opacity binning scheme, which is described in detail in \citet{N82,LJS94,VBS04}, is that  the model computations are significantly faster. This occurs because using even a relatively small number of opacity bins it is possible to maintain a sufficiently realistic heating/cooling balance in the model atmosphere. A question is, of course, how many bins should be deemed sufficient as this will obviously depend on the task at hand. We recall that the standard \COBOLD\ and \LHD\ models utilized in the present study were computed using five and six opacity bins at $\moh=0.0$ and $-2.0$, respectively. To test whether this number of bins indeed provides the required accuracy in the model computations, we also computed a test \COBOLD\ and \LHD\ models with $\Teff = 5000$\,K, $\log g = 2.5$, and $\moh = -2.0$ and the opacities grouped into 14 bins. Indeed, the differences in the magnitudes produced using the two models are minor (Fig.~\ref{fig:opacity-binning}), thus confirming the validity of the coarser binning scheme for the computation of SEDs, as well as the photometric magnitudes and colour indices.

\paragraph{Effects of the microturbulence velocity.}

\begin{figure}[tb]
        \centering
        \includegraphics[width=\columnwidth]{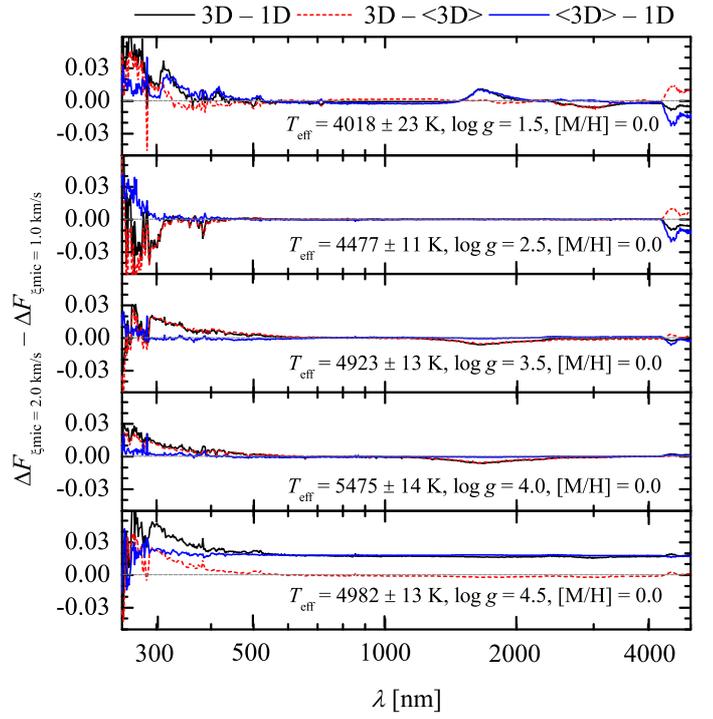}
        \caption
        {Effects of microturbulence velocity used calculating opacity distribution functions on the spectral energy distribution. The plot shows differences in the flux of several 3D hydrodynamical \COBOLD\ models computed using opacities with microturbulence velocities of 2.0 and 1.0~km/s.}
        \label{fig:micro}
\end{figure}

The standard ODFs from the \ATLAS\ package that were used to account for the line opacities in the computations of SEDs with the \NLTEIIID\ code were computed using the microturbulence velocity $\xi_{\rm mic}=2.0$\,km/s \citep{CK94}. In order to test how sensitive the computed SEDs and photometric magnitudes and colour indices were to this choice, we produced an additional set of SEDs using \ATLAS\ ODFs computed with $\xi_{\rm mic}=1.0$\,km/s. The obtained SEDs of the 3D, $\xtmean{\mbox{3D}}$, and 1D model atmospheres are shown in Fig.~\ref{fig:micro}. The differences between SEDs computed using the two values of microturbulence velocity are almost always below $\approx 1\%$, except in the blue part of the spectrum ($\lambda<400$\,nm) where in  extreme cases they may reach $\approx 4\%$ at $300$\,nm.

\paragraph{Open issues and future work.}

Results  obtained in our study show that spectral energy distributions -- and thus photometric magnitudes and colour indices -- are most sensitive to the influence of convection in the blue-UV part of the spectrum. The 3D--1D differences are largest there and tend to increase with decreasing wavelength. It is important to stress that at these wavelengths the influence of the chromospheric layers becomes increasingly important too.  For example, using our first exploratory 3D hydrodynamical \COBOLD\ model of a cool red giant in which the outer atmospheric layers were extended to mimic a chromosphere ($\Teff = 4000$\,K, $\log g = 1.5$, and $\moh = 0.0$), we found that the influence of chromosphere on the emergent model flux becomes significant bluewards of $\sim300$\,nm \citep{WKK17}. This is caused by additional heating in the outer chromosphere which leads to higher flux in the blue-UV. In fact, this influence is already noticeable at $1000$\,nm, but the differences start to exceed the level of $10$\% only at $\leq350$\,nm. Obviously, our exploratory model was still too simplistic to properly account for the physical complexity of real stellar chromospheres. Nevertheless, this test clearly demonstrated  that extending the realism of the current 3D hydrodynamical model atmospheres in this domain may be crucial for reliable emergent flux computations in the blue-UV part of the spectrum. 

It also unclear whether the numerical resolution of the model grid used in the 3D hydrodynamical simulations may also have an influence on the emergent SEDs, just as it does on the properties of spectral lines.

Finally, NLTE effects may also play a role, especially in the formation of emergent flux in the blue-UV part of the spectrum. Until now, however, NLTE effects could only be taken into account when modelling the formation of spectral lines of selected chemical elements with 3D hydrodynamical models. Therefore, the broader impact of these effects on the formation of continuum flux and line spectrum over a wider wavelength range is still unknown. Studies made using 1D hydrostatic model atmospheres have shown that there is a noticeable influence on the resulting SEDs if the line blanketing in the model atmospheres is computed in NLTE, in particular with the lower flux predicted in the UV part of the spectrum \citep[see e.g.][and references therein]{SCP12}. Due to large horizontal inhomogeneities in the 3D hydrodynamical models and additional heating in the lower chromosphere, the impact of NLTE effects may be expected to become even more significant, especially in the UV, and thus will not completely cancel out when computing the 3D--1D corrections. 

The bottom line therefore is that although the 3D--1D differences are largest in the blue-UV part of the spectrum, the SEDs in this wavelength range are also very sensitive to a variety of other effects, as discussed above. Therefore, one should be generally cautious with the predictions made using current generations of the model atmospheres in this wavelength range.

\section{Summary and conclusions\label{sect:conclus}}

We studied the influence of convection on the spectral energy distributions and photometric colour indices of stars accross the H--R diagram. For this, we used 3D hydrodynamical \COBOLD\ and 1D hydrostatic \LHD\ model atmospheres and computed their emergent wavelength-dependent fluxes (SEDs), which were further used to calculate magnitudes and colour indices in the Johnson-Cousins $UBVRI$, 2MASS $JHK_{\rm s}$, and Str\o{}mgren $uvby$ photometric systems. To assess the effect of the 3D--1D differences in the magnitudes and colour indices on stellar population studies, the model atmospheres were selected in such a way that their atmospheric parameters would cover a range occupied by stars in different Galactic populations, such as Galactic disk, bulge, and halo. 

The obtained results (Figs.~\ref{fig:col-dif_JCG}, \ref{fig:col-dif_Str}) show that the 3D--1D differences in the magnitudes and colour indices are generally small in both $UBVRIJHK_{\rm s}$ and $uvby$ photometric systems. Typically, the influence of convection is largest in the blue-UV part of the spectrum and is most strongly pronounced at lowest effective temperatures. The strongest effect is seen in cool red giants where the 3D--1D differences in the Johnson-Cousins $U$ and Str\o{}mgren $u$ bands may reach  $\approx-0.2$\,mag at $\moh=0.0$ and to $\approx-0.1$\,mag at $\moh=-2.0$. Noticeable differences are also seen in these bands in the MS and TO stars at $\moh=-2.0$,  although in this case they are of opposite sign and smaller, $\approx +0.05$\,mag. Nevertheless, in the majority of photometric bands the effect is significantly smaller and the 3D--1D differences typically do not exceed $\pm 0.03$\,mag. The effect of metallicity is strongest in the blue bands, but even in extreme cases the differences do not exceed $\approx 0.1$\,mag for magnitudes and colour indices computed at $\moh=0.0$ and $-2.0$. The influence of gravity is generally negligible in all bands studied.

While small, the 3D--1D differences in magnitudes and colour indices cannot be deemed unimportant. For example, even a relatively minor 3D--1D difference of $\Delta(V-I)=0.03$ in cool red giants is comparable to or larger than the typical uncertainty of photometric measurements and may lead to a difference in the effective temperature of $\approx 60$\,K. Moreover, the size of 3D--1D effects depends on the effective temperature, which will translate into different temperature corrections for cooler and hotter stars. 

Finally, we would like to stress that current 3D hydrodynamical models are obviously not free from different shortcomings and limitations. For example, scattering was treated as true absorption when computing the grid of 3D hydrodynamical \COBOLD\ model atmospheres that were used in this study. This may have an effect on the photometric magnitudes and colour indices, especially those located in the blue-UV part of the spectrum where the sensitivity to the effects of convection is most pronounced. Although our tests have shown that the impact of this and other modelling assumptions on the resulting 3D--1D differences in the photometric magnitudes and colour indices typically does not exceed a few hundredths of a magnitude, further improvements should certainly be expected with the arrival of next generation grids of 3D hydrodynamical model atmospheres.

\begin{acknowledgements}

We thank G.~Ambrazevi\v{c}i\={u}t\.{e}, V.~Dobrovolskas, and D.~Prakapavi\v{c}ius for their input during the various stages of this project. We also thank V.~Vansevi\v{c}ius for the useful comments and discussions. This research was funded by a grant (MIP-089/2015) from the Research Council of Lithuania. HGL and AK acknowledge financial support by the Sonderforschungsbereich SFB 881 ``The Milky Way System'' (subproject A4) of the German Research Foundation (DFG). HGL and MS acknowledge funding from the Research Council of Lithuania for the research visits to Vilnius.

\end{acknowledgements}

\bibliographystyle{aa}

\appendix
        
\section{A simple model of the colour corrections\label{sect_app:color-corr-model}}
        
To aid the interpretation of the calculated 3D--1D corrections of photometric magnitudes and colour indices, we developed a simple model that
qualitatively captures their behaviour. To this end, we envision the star as radiating like a black body. In the 3D case temperature fluctuations are present so that the radiative output stems from an ensemble of radiators of different temperature. The 1D case is described by a single temperature~$T$. The temperature fluctuations are characterized by their dispersion~\sig{T}. Since it is a rough model we derive the formulae to leading order in $\sig{T}/T$ only. As a normalization condition for the  ensemble of black-body radiators we obtain
        
                \newcommand{\aaa}{\ensuremath{\frac{2 h c^2}{\lambda^5}}}
                \newcommand{\bbb}{\ensuremath{\frac{hc}{\lambda k}}}
                \newcommand{\ebt}{\ensuremath{\exp{\left(\frac{hc}{\lambda kT}\right)}}}

                \beq
                \intlam \Blam(\lambda,T+\sig{T}) \approx \frac{\sigma}{\pi}T^4
                \left( 1+6\vartot \right) ,
                \label{e:norm}
                \eeq
                where $\sigma$ is the Stefan--Boltzmann constant, and $\Blam$ the Planck
                function given by
                \beq
                \Blam(\lambda,T) = \aaa \frac{1}{\ebt-1} 
                \label{e:planck}
                \eeq
                where $c$ is the speed of light, $k$ is the Boltzmann constant, and $h$ is the Planck constant. 
                The argument of $\Blam(\lambda,T+\sig{T})$ in Eq.~\eref{e:norm} is intended to
                indicate the expectation 
                value of the intensity emitted by the ensemble of black-body radiators.
                Expanding to the second order in $\dT$ (the linear order vanishes) we obtain
                \beq
                \frac{\Blam(T+\sig{T})}{\Blam(T)} \approx \left(1+ \frac{1}{2} T^2
                \frac{\Blam^{\prime\prime}(T)}{\Blam(T)}\vartot\right)\left(1+6\vartot\right)^{-1}
                \label{e:expand1}
                ,\eeq
                where the  primes indicate derivatives with respect to $T$. Evaluating the
                second derivative yields
                \beq
                \frac{\Blam^{\prime\prime}(T)}{\Blam(T)}\approx 
                \frac{\bbb\ebt\left[\ebt \left(\bbb-2T\right) +\bbb + 2T\right]}{T^4
                        \left(\ebt-1\right)^2} \,.
                \label{e:ddblam}
                \eeq
                
\begin{figure}
                        \centering
                        \includegraphics[angle=90,width=\columnwidth]{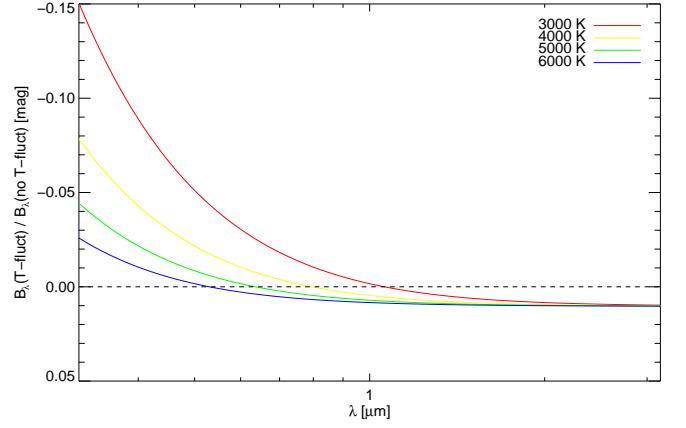}
                        \caption{Flux ratios as predicted the simple model given by
                                Eq.~\eref{e:expand1}.  For the plot we
                                assumed a temperature fluctuation of $\sig{T}/T=4\,\%$.}
                        \label{f:ddblam}
                \end{figure}

Figure~\ref{f:ddblam} depicts flux ratios as given by formula~\eref{e:expand1} for various values of the mean temperature $T$ assuming a fixed temperature dispersion of $\sig{T}/T=4\,\%$. The dispersion corresponds to an intensity contrast of roughly $16\,\%$ which is typical for the late-type stellar atmospheres studied here. From the figure some qualitative aspects of the colour corrections can be expected: i) the corrections become larger towards the blue part of the spectrum; ii) the corrections are greater for stars of low effective temperature; iii) the corrections have different signs in the blue and red. We emphasize that the model is only rough, and  leaves out the important response of the opacity to temperature fluctuations. Nevertheless, the detailed calculations show similarities to the predictions from the simple model.

\end{document}